\documentclass[12pt]{article}
\pdfoutput=1
\usepackage{wrapfig}
\usepackage{graphicx}
\usepackage{jheppub}
\usepackage{tikz}
\usepackage[T1]{fontenc}
\usetikzlibrary{arrows,snakes,backgrounds}

\title{$T$, $Q$ and periods in $SU(3)$ ${\cal N}=2$ SYM}

\author[a]{Davide Fioravanti,}
\author[a,b]{Hasmik Poghosyan,}
\author[b]{Rubik Poghossian}

\affiliation[a]{Sezione INFN di Bologna and
	Dipartimento di Fisica e Astronomia\\ Universita di Bologna,
	Via Irnerio 46, 40126 Bologna, Italy}
\affiliation[b]{Yerevan Physics Institute\\
	Alikhanian Br. 2, 0036 Yerevan, Armenia}
\emailAdd{fioravanti@bo.infn.it}
\emailAdd{hasmikpoghos@gmail.com}
\emailAdd{poghos@yerphi.am}

\abstract{We consider the third order differential equation derived from 
the deformed Seiberg-Witten differential for pure ${\cal N}=2$ SYM with gauge group $SU(3)$ in Nekrasov-Shatashvili limit of $\Omega$-background.
We show that this is the same 
differential equation that emerges in the context of Ordinary Differential Equation/Integrable 
Models (ODE/IM) correspondence for $2d$ $A_2$ Toda CFT with central charge $c=98$. 
 We derive the corresponding $QQ$ and related $TQ$ functional relations and establish 
 the asymptotic behaviour of $Q$ and $T$ functions at small instanton parameter $q \rightarrow 0$.
  Moreover, numerical integration of the Floquet monodromy matrix of the differential equation leads to evaluation of the $A$-cycles $a_{1,2,3}$ at any point of the moduli space of vacua parametrised by the vector multiplet scalar VEVs $\langle \textbf{tr}\,\phi^2\rangle$ and $\langle \textbf{tr}\,\phi^3\rangle$ 
even for large values of $q$ which are well beyond the reach of instanton calculus. The numerical results at small $q$ are in excellent agreement with instanton calculation. We conjecture  a very simple relation between Baxter's $T$-function and $A$-cycle periods $a_{1,2,3}$, which is an extension of Alexei Zamolodchikov's conjecture about Mathieu equation.
}

\begin{document}
\maketitle
\newcommand{\ie}{{\it i.e.\ }}
\def\bea{\begin{eqnarray}}
\def\eea{\end{eqnarray}}


\section*{Introduction}
Ever since 1994 when Seiberg and Witten  
derived exact low-energy Wilsonian effective action of (pure) $SU(2)$ ${\cal N}=2$ SYM \cite{Seiberg:1994rs},
 the interest in this kind of theories has been remaining extremely high. The reason is their remarkably rich physical and 
mathematical content. In fact, these theories provide a framework to address in a precise manner such 
problems as strong coupling, non-perturbative effects and confinement in non-Abelian   
gauge theory (so relevant, for instance, in the Standard Model). The impact of Seiberg-Witten theory 
in pure mathematics is also very substantial. Likely, the most famous applications are in algebraic geometry and topology of four-dimensional 
differentiable manifolds where e.g. the notions of Seiberg-Witten and Gromov-Witten invariants are 
of primary importance. 

The effective action of ${\cal N}=2$ SYM is given in terms of prepotential: a holomorphic 
function of the vacuum expectation values (VEV) of the vector multiplet scalar field. 
Large VEV expansion of the prepotential reveals its structure as sum of classical, one-loop 
and instanton contributions. Many researchers tried to restore the instanton contributions 
directly from the microscopic theory, but they succeeded only in the case of the first few instantons \cite{Dorey:2002ik}. Actual progress has been achieved with the idea of using equivariant 
localization techniques in the moduli space of instantons \cite{Flume:2001nc}, 
\cite{Flume:2001kb}, 
especially in combination 
with the introduction of the so-called $\Omega$ background (see \cite{Nekrasov:2002qd} and 
further developments \cite{Flume:2002az}, \cite{Nekrasov:2003rj}, \cite{Bruzzo:2002xf}). Considering theory in $\Omega$-background effectively embeds the system 
in a finite volume $\sim \frac{1}{\epsilon_1\epsilon_2}$, where the parameters 
$\epsilon_1$, $\epsilon_2$ are sort of 
angular velocities on orthogonal planes of (Euclidean) 4d space-time. This makes 
the partition function a finite, well defined quantity (commonly referred as Nekrasov 
partition function). Then the corresponding free energy coincides with (generalized) 
prepotential. The usual SW prepotential is recovered simply by sending the parameters 
$\epsilon_{1,2}\rightarrow 0$. Another, crucial consequence of introducing this background 
is the fact that instanton moduli integrals are localized at finitely many points. 
This property eventually leads to an elegant combinatorial formula for instanton 
contributions \cite{Flume:2002az}.

Later developments are even more surprising. It appears that introduction of $\Omega$ 
background is not merely a regularization trick. Thus,  keeping $\epsilon_{1,2}$ finite 
a deep relation between conformal blocks of 2d CFT and Nekrasov partition function 
\cite{Alday:2009aq}
emerges, so that the Virasoro central charge is related to these parameters, the masses of 
hypermultiplets specify inserted primary fields, while VEVs identify 
the states of the intermediate channel.

The special case  $\epsilon_{1}=-\epsilon_{2}$ bridges the theory with topological string, 
$\epsilon$-expansion of Nekrasov partition function coinciding with topological (string) genus expansion. 

Another special case of great interest is the Nekrasov-Shatashvili (NS) limit \cite{Nekrasov:2009rc}
when one of 
parameters, say $\epsilon_1$ is kept finite while $\epsilon_2=0$. From AGT point of 
view this case corresponds to semiclassical CFT when the central charge 
$c\rightarrow \infty$. Besides this, another interesting link to quantum integrable 
system emerges, now the remaining nonzero parameter $\epsilon_1$ being related to 
Plank constant. In NS limit many quantities familiar from original Seiberg-Witten 
theory become deformed or quantised in rather simple manner. 

In particular the algebraic equation defining 
Seiberg-Witten curve, becomes a finite difference equation \cite{Poghossian:2010pn}, 
which in terms of 
related integrable system is nothing but Baxter's $TQ$ equation (for a later development see also 
\cite{Bourgine:2017aqk}).
 Through discrete Fourier 
transform one gets a linear differential equation \cite{Fucito:2011pn}, which from 2d CFT 
perspective is 
the null vector decoupling equation \cite{Belavin:1984vu} in the semiclassical limit. 
This relation was an object of intensive investigations in the last decade (see e.g.    
\cite{Alday:2009fs,Mironov:2009uv,Maruyoshi:2010iu,Marshakov:2010fx,Nekrasov:2013xda,
Piatek:2011tp,Ashok:2015gfa,Poghossian:2016rzb,Poghosyan:2016mkh,
Nekrasov:2017gzb,Jeong:2017mfh}).

More recently, moving from Gaiotto's idea of looking at these
 equations as quantum versions of the (suitable power of the) SW differential \cite{Gaiotto:2009we}\footnote{In other words, the SW differential gives way to the oper upon quantisation.}, it has been proposed to investigate their monodromies (quantum periods over cycles) through the connection (Stokes) multipliers appearing in the  ODE/IM correspondence \cite{Dorey:1998pt,Bazhanov:1998wj}: \cite{Fioravanti:2019vxi} describes the general idea by exemplifying it in the simple case of pure $SU(2)$ gauge theory\footnote{Very few details are given for the cases with matter in the fundamental.} and in particular the link between the $a$-period and the Baxter transfer matrix $T$ function. In this perspective, Thermodynamic Bethe Ansatz  (TBA)-like considerations about pure $SU(2)$ gauge theory were initiated in \cite{Gaiotto:2014bza} at zero modulus (of the Coulomb branch) for the dual period $a_D$, and then more recently pursued by \cite{Grassi:2019coc}.

In fact, in this paper we show how to compute the gauge A-periods of the pure $SU(3)$ theory (without any matter hypermultiplet: {\it cf.} \cite{Klemm:1994qs} and \cite{Argyres:1994xh} for what concerns the generalisation of SW theory to higher rank gauge groups) as Floquet monodromy coefficients of the aforementioned differential equation (in the complex domain). Then, we propose a connexion between them and the integrable Baxter's $T$ function which extends non trivially what happens in the $SU(2)$ case and shows that the latter is not an accident. More in details, we obtain a third order linear differential equation with some similarities (and differences\footnote{The main relevant difference is, as in simpler $SU(2)$ case \cite{Zamolodchikov:2000unpb} and \cite{Fioravanti:2019vxi}, the exit of the oper parameter ($M>1/2$ in \cite{Dorey:1998pt}) from the range of validity with the appearance of a extra irregular singularity in zero (besides that at $\infty$).}) with the third order oper of  ODE/IM correspondence in  
\cite{Dorey:1999pv}, \cite{Bazhanov:2001xm}. As the latter correspond to some 'minimal' case $M>1/2$, we may conjecture, along the lines of \cite{Dorey:1999uk} and \cite{Zamolodchikov:2000unpb}, that we are describing $A_2$-Toda CFT with central charge $c=98$.

In a very interesting unfinished paper \cite{Zamolodchikov:2000unpb} Alexei Zamolodchikov has proposed ODE/IM for the Liouville CFT TBA. Special attention has been payed to the self-dual case $c=25$, when the related ODE becomes the {\it modified} Mathieu equation and a elegant 
relationship between Floquet exponent and Baxter's $T$ function has been suggested. As written, the implication of this conjecture for the period on the $A$-cycle of (effective) $SU(2)$ gauge theory has been highlighted and used by \cite{Fioravanti:2019vxi}. But it was not clear from there if and how it is possible to generalise this beautiful connection between transfer matrix and periods for higher rank groups.  

In the details of this paper we derive $QQ$ and $TQ$ functional relations (see eqs. (\ref{Qc_QQ}), (\ref{T_Q})) and extend Zamolodchikov's 
conjecture for the case of gauge group $SU(3)$ (see eq. (\ref{conj_su3})). We show that numerical integration of the differential equation leads to 
evaluation of the 'quantum' $A$-cycle periods $a_{1,2,3}$ at any point of the moduli 
space of vacua parametrised by the vector multiplet scalar VEV's 
$u_2=\langle \textbf{tr}\,\phi^2\rangle$ and $u_3=\langle \textbf{tr}\,\phi^3\rangle$ 
even for large values of $q$ at which the instanton series diverges.    
We have checked that the numerical results at small $q$ are in excellent agreement with instanton 
calculation. Thus the main message of this paper is that the differential equation provides 
an excellent tool for investigation of deformed SW theory in its entire range 
from weak to strong coupling.    

The paper is organized as follows:

Section \ref{NPF} is a short review on instanton calculus for $SU(N)$ SW theory without
hypers in $\Omega$-background. 
Here one can find explicit expressions as a sum over (multiple) Young diagrams for 
Nekrasov partition function and VEV's  $\langle \textbf{tr}\,\phi^J\rangle$.
 
Section \ref{BDE} is a brief introduction to deformed SW theory. 
We present the main results of \cite{Poghossian:2010pn}  in a form convenient 
for our present purposes.  
Starting from section \ref{DE} we consider the case of $SU(3)$ theory. The main 
tool of our investigation, a third order linear ODE is derived and its asymptotic 
solutions are found.

In section \ref{FR} we identify a unique solution $\chi(x)$ which rapidly vanishes for 
large negative values of the argument $x\rightarrow -\infty $. The three quantities 
$Q_{1,2,2}$ are defined as coefficients of expansion of $\chi(x)$ in terms of 
three independent solutions $U_{1,2,3}(x)$ defined in asymptotic region $x\gg 0$. 
Investigating symmetries of the differential equation we find a system of difference 
equations for $Q_k$ and their analogs $\bar{Q}_k$ obtained by flipping the sign of
parameter $u_3\rightarrow -u_3$. Based on this $QQ$ system we introduce Baxter's 
$T$ function and write down corresponding $TQ$ relations. 
    
In section \ref{a_cycles} we show how numerical integration of the 
differential equation along imaginary direction with standard boundary 
conditions allows one to find the monodromy matrix and corresponding 
Floquet exponents, which in the context of gauge theory, coincide with the $A$-cycle 
periods $a_{1,2,3}$. We have convincingly demonstrated the correctness of this identities
 trough comparison with instanton computation. But the main value of this method is 
 that it makes accessible also the region of large coupling constants, which is beyond 
 the reach of instanton calculus. Eventually, we close this section by suggesting a simple relation between Baxter's $T$-function and 
$A$-cycle periods $a_{1,2,3}$ of $SU(3)$ theory, which can be thought of as a natural extension of Alexei Zamolodchikov's 
conjecture relating Floquet exponent of Mathieu equation to Baxter's $T$ 
function in $c=25$ Liouville CFT.

Finally appendix \ref{TQ_proof} contains few technical details 
for derivation of the $TQ$ relation.

\section{Nekrasov partition function and the VEVs $\langle \textbf{tr}\,\phi^J\rangle $}
\label{NPF}
Consider pure $SU(N)$ theory without hypers in $\Omega$-background.
The instanton part of partition function is given by \cite{Nekrasov:2002qd}
\bea 
Z_{inst}(\mathbf{a},\epsilon_1,\epsilon_2,q)=\sum_{\vec{Y}}Z_{\vec{Y}} \left((-)^Nq\right)^{|\vec{Y}|}
\,,
\label{z_inst}
\eea
where sum runs over all $N$-tuples of Young diagrams $\vec{Y}=(Y_1,\cdots,Y_N)$ 
, $|\vec{Y}|$ is the total number all boxes,
$\mathbf{a}=(a_1,a_2,\cdots,a_N)$ are VEV's of adjoint scalar from 
${\cal N}=2$ vector multiplet, $\epsilon_1$, $\epsilon_2$, as already mentioned, 
parametrize the $\Omega$-background  and the instanton counting parameter $q=\exp 2\pi i 
\tau $, with  $\tau=\frac{i}{g^2}+\frac{\theta}{2\pi}$
being the (complexified) coupling constant. The coefficients $Z_{\vec{Y}}$ are factorized 
as
\bea 
\label{ZY_factorization}
Z_{\vec{Y}}=\prod_{u,v=1}^N\frac{1}{P(Y_u,a_u|Y_v,a_v)}\,\,,
\eea 
where the factors $P(\lambda,a|\mu,b)$ for 
arbitrary pair of Young diagrams $\lambda,\mu$ and associated VEV parameters $a$, $b$, 
are given explicitly by the formula \cite{Flume:2002az} 
\bea
\label{P_factor} 
&&P(\lambda ,a|\mu ,b)=\\
&&\quad \prod_{s\in \lambda}(a-b+\epsilon_1(1+L_\mu(s))-\epsilon_2 A_\lambda(s))
\prod_{s\in \mu}(a-b-\epsilon_1L_\lambda(s)+(1+\epsilon_2 A_\lambda(s)))\nonumber
\eea
If one specifies location of a box $s$ by its horizontal and vertical 
coordinates $(i,j)$, so that $(1,1)$ corresponds to the corner box, its
leg length $L_\lambda(s)$ and arm length $A_\lambda(s)$ with respect to the 
diagram $\lambda$ ($s$ does not necessarily belong to $\lambda$) are defined as
\bea 
A_\lambda(s)=\lambda_i-j; \qquad\qquad L_\lambda(s)=\lambda_j'-i\,,
\eea
where $\lambda_i$ ($\lambda_j'$) is $i$-th column ($j$-th row) of diagram $\lambda$ 
with convention that when $i$ exceeds the number of columns 
($j$ exceeds the number of rows) of $\lambda$, one simply sets $\lambda_i=0$
($\lambda_j'=0$).
The instanton part of (deformed) prepotential is given by \cite{Nekrasov:2002qd}
\bea 
F_{inst}({\bf a},q)=-\epsilon_1\epsilon_2\log Z_{inst}\,.
\eea
Instanton calculus allows one to obtain also the VEV's $\langle \textbf{tr}\,\phi^J\rangle$, 
$\phi$ being the adjoint scalar of vector multiplet:
\bea
\langle \textbf{tr}\,\phi^J\rangle=\sum_{i=1}^Na_u^J+
Z_{inst}^{-1}\sum_{\vec{Y}}Z_{\vec{Y}}{\mathcal O}_{\vec{Y}}^J q^{|\vec{Y}|}
\,,
\label{tr_phiJ}
\eea 
where $Z_{\vec{Y}}$ is already defined by (\ref{ZY_factorization}), (\ref{P_factor}), 
and \cite{Losev:2003py,Flume:2004rp}
\bea 
{\mathcal O}_{\vec{Y}}^J=\sum_{u=1}^N\,\sum_{(i,j)\in Y_u}
\left(\left(a_u+\epsilon_1 i+\epsilon_2(j-1)\right)^J+
\left(a_u+\epsilon_1 (i-1)+\epsilon_2j\right)^J\right.\nonumber\\
-\left.\left(a_u+\epsilon_1 (i-1)+\epsilon_2(j-1)\right)^J
-\left(a_u+\epsilon_1 i+\epsilon_2 j\right)^J\right). 
\eea   
\section{A Baxter difference equation}
\label{BDE}
\subsection{Bethe ansatz equation for NS limit}
It was shown in \cite{Poghossian:2010pn} that in NS limit $\epsilon_2\rightarrow 0$, the sum 
(\ref{z_inst}) is dominated by a single term corresponding to a unique 
array of Young diagrams $\vec{Y}^{(cr)}$ specified by properties 
(the $i$-th column length of a diagram $Y_u$ will be denoted as $Y_{u,i}$):
\begin{itemize}
\item{Though the total number of boxes $\rightarrow \infty $ in 
$\epsilon_2\rightarrow 0$ limit the rescaled column lengths 
$\epsilon_2 Y^{(cr)}_{u,i}$, converge to finite values 
\[
\xi_{u,i}=\lim_{\epsilon_2\rightarrow 0}\epsilon_2 Y^{(cr)}_{u,i}\,.
\]
}
\item{
The rescaled column lengths  at small $q$ behave as $\xi_{u,i}\sim O(q^i)$. 
This means in particular, that in order to achieve accuracy up to $q^L$, 
it is consistent to consider restricted Young diagrams with number of columns $\le L$. 
}
\item{
Up to arbitrary order $q^L$ the quantities 
\[
x_{u,i}=a_u+\epsilon_1(i-1)+\xi_{u,i}
\]
satisfy the Bethe-ansatz equations (for each $u=1,2,\cdots N$)
\bea
-q \prod_{v,j}^{N,L} \frac{(x_{u,i}-x_{v,j}-\epsilon_1)(x_{u,i}-x_{v,j}^0+ 
	\epsilon_1)}{(x_{u,i}-x_{v,j}+\epsilon_1)(x_{u,i}-x_{v,j}^0-\epsilon_1)}
=\prod_{v=1}^N(x_{u,i}-a_v+\epsilon_1)(a_v-x_{u,i})\,,\quad
\label{BA}
\eea
where, by definition
\[
x_{u,i}^0=a_u+\epsilon_1(i-1)\,.
\]
}	
\end{itemize}
The system of equations (\ref{BA}) together with the property $\xi_{u,i}\sim O(q^i)$ 
uniquely fixes the quantities $x_{u,i}$ up to order $q^L$. Of course, calculations become 
more cumbersome if one increases $L$. Examples of explicit computations for first few 
values of $L$ can be found in \cite{Poghossian:2010pn}. 
\subsection{Baxter's difference equation and deformed Seiberg-Witten 'curve'}
The BA equations can be transformed into a difference equation  \cite{Poghossian:2010pn}
\bea
Y(z+\epsilon_1)+\frac{q}{\epsilon_1^{2N}}Y(z-\epsilon_1)=
\epsilon_1^{-N} P_N(z+\epsilon_1)\, Y(z)\,,
\label{difference_eq}
\eea 
where $Y(z)$ is an entire function with zeros located at $z=x_{u,i}$:
\bea
Y(z)=\prod_{u=1}^N e^{\frac{z}{\epsilon_1} \psi(\frac{a_u}{\epsilon_1})}\prod_{i=1}^\infty 
\left(1-\frac{z}{x_{u,i}}\right)e^{z/x_{u,i}^0}\,,
\label{Y}
\eea
and 
\[
\psi(x)=\frac{d}{dx}\log \Gamma(x)
\]
is the logarithmic derivative of Gauss' gamma-function. Finally $P_N(z)$ is 
an $N$-th order polynomial which parametrises the Coulomb branch of the theory. 
Explicit expressions of coefficients of this polynomial in terms of VEVs 
\bea
u_J\equiv \langle {\textbf tr} \phi^J \rangle
\label{u_J_def}
\eea  
will be presented later for the case of our current interest $N=3$. For more 
general cases one can refer to \cite{Poghossian:2010pn}. Now, let us briefly recall how the 
difference equation (\ref{difference_eq}) is related to the Seiberg-Witten curve.
Introducing the function 
\[
y(z)=\epsilon_1^{N}\,\,\,
\frac{Y(z)}{Y(z-\epsilon_1)}
\]
one can rewrite (\ref{difference_eq}) as
\bea
y(z)+\frac{q}{y(z-\epsilon_1)}=P_N(z)\,.
\label{DSW} 
\eea
At large $z$ the function $y(z)$ behaves as
\[
y(z)=z^N(1+O(1/z))\,.
\] 
Notice that setting $\epsilon_1=0$ in (\ref{DSW}) one obtains 
an equation of hyperelliptic curve, which is just the Seiberg-Witten curve. 
When $\epsilon_1\neq 0$, everything goes surprisingly similar to the original Seiberg-Witten theory. For example the r\^ole of Seiberg-Witten 
differential is played anew by the quantity 
\[
\lambda_{SW}=z \frac{d}{dz}\log y(z) \, ,
\]  
and, as in the undeformed theory, the expectation values (\ref{u_J_def}) are given by the contour integrals
\bea
\langle \textbf{tr}\,\phi^J\rangle =\oint_{\cal C} \frac{dz}{2\pi i} z^J\partial_z \log y(z)\,,
\label{u_J_int}
\eea
where $\cal C$ is a large contour, enclosing all zeros and poles of $y(z)$.
\subsection{Details on $SU(3)$ theory}
Without any essential loss of generality, from now on we will assume that 
\bea
u_1 \equiv  \langle\textbf{tr}\,\phi \rangle=a_1+a_2+a_3=0\,.
\eea
Representing $y(z)$ as a power series in $1/z$
\bea
y(z)=z^3(1+c_1z^{-1}+c_2z^{-2}+c_3z^{-3}+\cdots )
\label{y_expansion}
\eea 
and inserting in eq. (\ref{u_J_int}) one easily finds the relations 
\bea 
c_1=0; \qquad c_2=-\frac{u_2}{2}; \qquad c_3=-\frac{u_3}{3}\,.
\label{c_u}
\eea 
Now, consistency of (\ref{y_expansion}), (\ref{c_u}) and (\ref{DSW}) 
immediately specifies the polynomial $P_3(z)$ (we omit the 
subscript $3$, since only the case $N=3$ will be considered later on)
\bea 
P(z)=z^3-\frac{u_2}{2}\,z-\frac{u_3}{3}\,\,.
\eea
\section{The differential equation and its asymptotic solutions}
\label{DE}
\subsection{Derivation of the differential equation}
To keep expressions simple, from now on we will set $\epsilon_1=1$. 
In fact, at any stage the $\epsilon_1$ dependence can be easily restored on dimensional grounds. Taking the results of previous subsection, the difference equation
for $N=3$ case (\ref{difference_eq}) can be rewritten as
\bea
Y(z)-\left(z^3-\frac{u_2}{2}z-\frac{u_3}{3}\right) Y(z-1)+q\,Y(z-2)=0
\,,
\label{difference_eq_3}
\eea
By means of inverse Fourier transform, following \cite{Fucito:2011pn,
Nekrasov:2013xda,Poghossian:2016rzb}, from (\ref{difference_eq_3}) we can derive a third order linear differential equation for the function
\bea 
f(x)=\sum_{z\in\mathbb{Z}+a}e^{x (z+1)}Y(z)\,.
\label{f_series}
\eea
At least when  $|q|$ is sufficiently small, it is expected that the series 
is convergent for finite $x$, 
provided $a$ takes one of the three possible values $a_1$, $a_2$ or $a_3$.  
Taking into account the difference relation (\ref{difference_eq_3}), one 
can easily check that  the function (\ref{f_series}) solves the 
differential equation
\bea 
-f^{\,'''}(x)+\frac{u_2}{2}\,f^{\,'}(x)+\left(e^{-x}+q\,e^x+\frac{u_3}{3}\right)
f(x)=0\,.
\label{diff_eq1}
\eea 
Denoting
\[ 
q=\Lambda^6 
\]
and shifting the variable
\[
x\rightarrow x-\log \Lambda^3
\] 
the differential equation (\ref{diff_eq1}) may be cast into a more symmetric 
form
\bea 
-f^{\,'''}(x)+\frac{u_2}{2}\,f^{\,'}(x)+\left(\Lambda^3(e^x+e^{-x})+\frac{u_3}{3}\right)
f(x)=0\,.
\label{diff_eq2}
\eea
\subsection{Solutions at $x\rightarrow \pm \infty $}
Physics leads us to introduce parameters $p_1$, $p_2$, $p_3$ 
satisfying $p_1+p_2+p_3=0$ such that
\bea
u_2=p_1^2+p_2^2+p_3^2=2(p_1^2+p_2^2+p_1p_2); \quad 
u_3=p_1^3+p_2^3+p_3^3=-3p_1 p_2 (p_1+p_2)\,, \qquad
\label{u_2_3}
\eea
as in the weak coupling limit $\Lambda \rightarrow 0$ the parameters $p_i$ and $a_i$, respectively, coincide. 

At large positive values $x\gg 3\ln \Lambda$ the term $\Lambda^3 e^{-x}$ in (\ref{diff_eq2}) can be neglected. In this region the differential equation 
can be solved in terms of hypergeometric function $_0F_2(a,b;z)$ 
defined by the power series
\bea 
_0F_2(a,b;z)=\sum_{k=0}^{\infty}\frac{z^k}{k!(a)_k(b)_k}\,,
\eea
where 
\bea
(x)_k=x(x+1)\cdots(x+k-1)
\eea
is the Pochhammer symbol. Three linearly independent solutions can be chosen as
\bea
U_i(x)\approx e^{(x+3\theta )p_i}\,\,
_0F_2(1+p_i-p_j,1+p_i-p_k;e^{x+3\theta })\,\,,
\label{U_solutions}
\eea 
where by definition 
\bea
\Lambda \equiv \exp \theta
\eea
and the indices $(i,j,k)$ 
are cyclic permutations of $(1,2,3)$. We used the symbol $\approx$ 
in (\ref{U_solutions}) to mean that the approximations of the solutions hold, striktly speaking, only for $x\gg 3\theta $ (at leading order). In the end, we must verify that the Wronskian of the three solutions (\ref{U_solutions}) (below and later on, for brevity, we use the notation $p_{ij}\equiv p_i-p_j$)
\bea
Wr[U_1(x),U_2(x),U_3(x)]\equiv \det \left( 
\begin{array}{ccc}
U_1(x)&U_2(x)&U_3(x)\\
U_1^{'}(x)&U_2^{'}(x)&U_3^{'}(x)\\
U_1^{''}(x)&U_2^{''}(x)&U_3^{''}(x)
\end{array}
\right) 
=p_{12}p_{23}p_{31}\quad 
\label{U_wr}
\eea
is not zero provided the parameters $p_i$ are pairwise different. Thus, (\ref{U_wr}) confirms that generically the $U_i(x)$ are linearly independent 
and constitute a basis in the space of all solutions. 

Similarly in region $x\ll -3\theta $ the term $\Lambda^3 e^{x}$ 
of (\ref{diff_eq2}) becomes negligible and one can write down the three linear independent solutions 
\bea
V_i(x)\approx e^{(x-3\theta )p_i}\,\,
_0F_2(1-p_i+p_j,1-p_i+p_k;-e^{-x+3\theta })\,\,.
\label{V_solutions}
\eea
In fact, we obtain the same result for the Wronskian \bea
Wr[V_1(x),V_2(x),V_3(x)] =p_{12}p_{23}p_{31} \, .
\label{V_wr}
\eea
\section{The functional relations}
\label{FR}
\subsection{The $QQ$ relations}
All three solutions $V_i(x)$ grow very fast at $x\rightarrow -\infty$, but 
there is a special linear combination (unique, up to a common constant factor) which vanishes in this limit. If it is the fastest one (as we suspect), this solution is usually referred to as subdominant. Using formulae for asymptotics of $_0F_2$, which 
can be found e.g. in \cite{NIST:DLMF}, we are able to establish that the correct combination is
 \bea
\label{chi}
&\chi(x)=
\frac{ \Gamma (p_{12}) \Gamma ( p_{13}) }{4 \pi ^2}\, V_1(x)+
\frac{ \Gamma (p_{23}) \Gamma ( p_{21}) }{4 \pi ^2}\, V_2(x)+
\frac{ \Gamma (p_{31}) \Gamma ( p_{32}) }{4 \pi ^2}\, V_3(x)\,.
\eea
Its asymptotic expansion at $x\rightarrow -\infty$ is given by
 \bea
\chi (x)=
\frac{v^{-\frac{1}{3}} e^{-3 v^{1/3}} }{2 \pi  \sqrt{3}}
\left(
1-\left(\frac{1}{9}-\frac{u_2}{2}\right) v^{-\frac{1}{3}}+ 
\left(\frac{u_2^2}{8}-\frac{5 u_2}{36}+\frac{u_3}{6}+\frac{2}{81}\right)v^{-\frac{2 }{3}} \right.\nonumber\\-
\left. \left(-\frac{u_2^3}{48}+\frac{u_2^2}{18}-\frac{u_3 u_2}{12}-\frac{13 u_2}{324}+\frac{7 u_3}{54}+\frac{14}{2187}\right)v^{-1} +O\left(v^{-\frac{4}{3}}\right)\right)\,,\quad 
\eea
where we denoted 
\[
v=\exp (3\theta-x)
\]
and $u_2$, $u_3$ are defined in terms of $p_i$ in (\ref{u_2_3}).

Since $U_i(x)$ constitute a complete set of solutions one can represent $\chi(x)$ as 
a linear combination
\bea
\label{chiU}
\chi (x,\theta)=\sum_{n=1}^{3}Q_n(\theta)
\Gamma(p_{nj})\Gamma(p_{nk}) e^{-3p_n\theta}U_n(x,\theta),
\eea
where the important quantities $Q_n(\theta)$, based on general theory of linear 
differential equations, are expected to be entire functions of $\theta$ (and also of
parameters ${\bf p}$ dependence on which will be displayed explicitly only if necessary).  
The following, easy to check property plays an essential role in further discussion. 
Namely the Wronskian of any two solutions $f(x)$, $g(x)$ 
of the differential equation (\ref{diff_eq2}) 
\[
W[f(x),g(x)]\equiv f(x)g'(x)-g(x)f'(x)
\] 
satisfies the {\it adjoint} equation, i.e. the one obtained by reversing the signs $\bf{p}\rightarrow -\bf{p}$ 
and $\Lambda^3\rightarrow -\Lambda^3$. Taking inspiration from this property, it is then possible to show exactly that
\bea
\label{Wr_chi_chi}
Wr\left[\chi (x,\theta+\frac{i \pi}{3}),\chi (x,\theta-\frac{i \pi}{3})\right]=
-\frac{i}{2 \pi }\bar{\chi}(x,\theta)\,,
\eea
where $\bar{\chi}(\theta)=\chi(\theta,-\mathbf{p})$. In fact, the property entails that the l.h.s. of (\ref{Wr_chi_chi}) satisfies the differential equation 
(\ref{diff_eq2}) with substitution $\bf{p}\rightarrow -\bf{p}$.  Besides, by using the identity\footnote{It can be proven, for instance, by expanding both sides in powers of $e^{-x}$.}
\bea
&&Wr\left[e^{\frac{2-a-b}{3}\,x}\,_0F_2(a,b,-e^{-x})\,,e^{\frac{2a-b-1}{3}\,x}\,
_0F_2(2-a,1-a+b,-e^{-x})\right]\nonumber\\
&&=(a-1)e^{\frac{1+a-2b}{3}\,x}\,_0F_2(b,1-a+b,e^{-x})\,, 
\label{Wr_F_F}
\eea
it is not difficult to show the match of the $x\rightarrow-\infty$ asymptotics of both sides. Of course, the combination of these two statements implies the equality (\ref{Wr_chi_chi}) everywhere.

Let us investigate the $x\rightarrow \infty$ limit of (\ref{Wr_chi_chi}). Taking into account 
(\ref{chiU}) and using the identity (\ref{Wr_F_F}) (with $x$ substituted by $-x$), 
we obtain the functional relations
\bea
\label{Qc_QQ}
\frac{\sin(\pi p_{jk})}{2i \pi^2}\, \bar{Q}_n(\theta)=
Q_j\left(\theta+\frac{i \pi}{3}\right)Q_k\left(\theta-\frac{i \pi}{3}\right)-
Q_j\left(\theta-\frac{i \pi}{3}\right)Q_k\left(\theta+\frac{i \pi}{3}\right),\qquad 
\eea
where again, the bar on $Q_n$ indicates the sign change $\bf{p}\rightarrow -\bf{p}$
\[
\bar{Q}_n(\theta,{\bf p})\equiv Q_n(\theta,-\bf{p})
\] 
and  $(n,j,k)$ is a permutations of $(1,2,3)$.

At the end of this section let us establish the $\theta\rightarrow-\infty $ asymptotics 
of $Q_k(\theta)$ and  $\bar{Q}_k(\theta)$. Obviously, in this case both (\ref{U_solutions})
and  (\ref{V_solutions}) are approximate solutions of (\ref{diff_eq2}) at $x\sim 0$. 
Thus, comparison of (\ref{chi}) with (\ref{chiU}) ensures that for $\theta\ll 0$
\bea 
\label{Q_asymp}
Q_k(\theta)\sim \frac{\exp (-3\theta p_k)}{4 \pi^2}\,;\qquad 
\bar{Q}_k(\theta)\sim \frac{\exp (3\theta p_k)}{4 \pi^2}\,.
\eea
It is easy to see that above asymptotic behavior is fully  consistent
with functional relations (\ref{Qc_QQ}). 
\subsection{$SU(3)$ version of Baxter's $TQ$ relation}
The functional relations (\ref{Qc_QQ}) suggest the following $SU(3)$ analog 
of Baxter's $TQ$ equations:
\bea
\label{T_Q}
&T(\theta)Q_j\left(\theta-\frac{\pi i}{6}\right)\bar{Q}_k\left(\theta+\frac{\pi i}{6}\right)  
=\hspace{9.6cm}\\
&Q_j(\theta-\frac{5\pi i}{6})\bar{Q}_k\left(\theta+\frac{\pi i}{6}\right)
+Q_j\left(\theta+\frac{\pi i}{2}\right)\bar{Q}_k\left(\theta-\frac{\pi i}{2}\right)+
Q_j\left(\theta-\frac{\pi i}{6}\right)\bar{Q}_k\left(\theta+\frac{5\pi i}{6}\right)\nonumber  
\eea 
for $j,k\in \{1,2,3\}$ with $j\neq k$. To uncover the essence of this construction, notice 
that for a fixed pair of indices $(i,j)$ (\ref{T_Q}) can be thought as definition 
of function $T(\theta)$ in terms of $Q$'s. Then the nontrivial question is ``do 
other choices of $(j,k)$ lead to the same $T$?" Fortunately, elementary algebraic manipulations 
with the help of (\ref{Qc_QQ}) ensure that the answer is positive. As mentioned 
earlier, $Q_i(\theta)$ are  entire functions. A thorough analysis shows that due to  
(\ref{Qc_QQ}) all potential poles of $T(\theta)$ have zero residue. Thus $T(\theta)$ 
is an entire function too.  Details on proofs of above two statements can be 
found in appendix \ref{TQ_proof}.
The Bethe ansatz equations can be represented as (see equality (\ref{T_entire}))
\bea 
\frac{Q_j(\theta_\ell-\frac{2\pi i}{3})\bar{Q}_k\left(\theta_\ell+\frac{\pi i}{3}\right)}{
Q_j\left(\theta_\ell+\frac{2\pi i}{3}\right)\bar{Q}_k\left(\theta_\ell-\frac{\pi i}{3}\right)}=-1\,,
\eea 
where $\theta_\ell$ are the zeroes of $Q_j(\theta)$.  

Functional relations similar to (\ref{Qc_QQ}) and (\ref{T_Q}) emerge also in the context of ODE/IM for 'minimal' 2d CFT with extra spin $3$ current ($W_3$ symmetry) \cite{Dorey:1999pv}, \cite{Bazhanov:2001xm}. From there we can extrapolate that our case might correspond to the special choice 
of Virasoro central charge $c=98$ for Toda CFT. In fact, this value of the central charge lies outside the region discussed in above references. Nevertheless, it should be possible to derive the corresponding TBA equations: we leave this task for future publication. 
\section{Quantum periods and prepotential from Floquet monodromies and extension of Zamolodchikov's conjecture}
\label{a_cycles}
\subsection{The Floquet-Bloch monodromy matrix}
\label{sect_MM}
Consider the basis of solutions $f_1(x)$, $f_2(x)$, $f_3(x)$ of (\ref{diff_eq2}) with standard 
initial conditions 
($n,k\in\{1,2,3\}$)
\bea
\label{BC}
\left.f^{(k-1)}_n(x)\right|_{x=0} =\delta_{k,n} \, .
\eea
Since the functions $f_n(x+2\pi i)$ are solutions too, we can
define the monodromy matrix $M_{k,n}$ as
\bea 
f_n(x+2\pi i)=\sum_{k=1}^3 f_k(x)M_{k,n}
\eea
Clearly
\bea
\label{M_matrix}
M_{k,n}=f_n^{(k-1)}(2\pi i)\nonumber \, .
\eea
The solutions (\ref{f_series}) with $a\in \{a_1,a_2,a_3\}$ have diagonal monodromies 
and can be represented as certain linear combinations of $f_n(x)$. 
In other words the eigenvalues of the monodromy matrix $M_{k,n}$ must be identified with $\exp (2\pi ia_k)$, with $k=1,2,3$:
\bea 
\label{spec_M}
Spec(M_{k,n})=\{\exp (2\pi ia_1),\exp (2\pi ia_2),\exp (2\pi ia_3)\} \, .
\eea 

For any fixed values of parameters $\Lambda$, ${\bf p}$, it is easy to integrate 
numerically the differential equation (\ref{diff_eq2}) with boundary conditions 
(\ref{BC}), find the matrix $M_{k,n}$ and then its eigenvalues $\exp (2\pi ia_n)$. Taking into account Matone relation \cite{Matone:1995rx}, valid also in the presence of $\Omega$-background \cite{Flume:2004rp}, 
\bea
u_2\equiv \langle \textbf{tr}\,\phi^2\rangle=\sum_{n=1}^3a_n^2+2q\partial_qF_{inst}(q,{\bf a}) \, ,
\eea   
we can access the deformed prepotential for any value of the coupling constant.
\subsection{Comparison of the instanton counting against numerical results}
Using formula of section \ref{NPF} it is straightforward to  calculate 
$\langle \textbf{tr}\,\phi^2\rangle$ or $\langle \textbf{tr}\,\phi^3\rangle$ as a 
power series in $q$.
Here are the 3-instanton results (it is assumed that $a_1+a_2+a_3=0 $ and 
by definition $a_{jk}\equiv a_j-a_k$) 
\bea 
\label{tr_phi2}
\langle \textbf{tr}\,\phi^2\rangle &=&\sum_{k=1}^3a_k^2
-\frac{12(1-h_2)q}{\prod_{j<k}(a_{jk}^2-1)}
+\frac{P_{2,2}q^2}{\prod_{j<k}(a_{jk}^2-1)^3(a_{jk}^2-4)}+O(q)^4\qquad \\ 
\label{tr_phi3}
\langle \textbf{tr}\,\phi^3\rangle &=&\sum_{k=1}^3a_k^3
+\frac{54 h_3q}{\prod_{j<k}(a_{jk}^2-1)}
+\frac{P_{3,2}q^2}{\prod_{j<k}(a_{jk}^2-1)^3(a_{jk}^2-4)}\nonumber\\
&-&\frac{P_{3,3}q^3}{\prod_{j<k}(a_{jk}^2-1)^5(a_{jk}^2-4)(a_{jk}^2-9)}
+O(q)^4\,,
\eea  
where
\bea 
h_2=\frac{a_1^2+a_2^2+a_3^2}{2}\,;\qquad h_3=-a_1a_2a_3\,,
\eea
and 
\bea 
P_{2,2}=36 (220 - 1027 h_2 + 1659 h_2^2 - 698 h_2^3 - 958 h_2^4 + 1257 h_2^5 - 
521 h_2^6\hspace{3cm}\\
+ 68 h_2^7 - 13959 h_3^2 + 33804 h_2 h_3^2 - 
25434 h_2^2 h_3^2 + 5292 h_2^3 h_3^2\hspace{2.8cm}\nonumber\\
+ 297 h_2^4 h_3^2 + 13851 h_3^4 - 
5103 h_2 h_3^4)\hspace{3.5cm}\nonumber\\
P_{3,2}=-162 h_3 (455 - 2487 h_2 + 4602 h_2^2 - 3286 h_2^3 + 291 h_2^4 + 
573 h_2^5 - 148 h_2^6\hspace{2cm}\\
- 8073 h_3^2 + 14985 h_2 h_3^2 - 7695 h_2^2 h_3^2 + 
783 h_2^3 h_3^2 + 1458 h_3^4)\hspace{1.8cm}\nonumber\\
P_{3,3}=-108 h_3 (12078563 - 109310145 h_2 + 400164948 h_2^2 - 722480972 h_2^3
\hspace{2.5cm}\\ + 
538752402 h_2^4 + 275687658 h_2^5 - 946955868 h_2^6 + 
865056708 h_2^7\hspace{3.1cm}\nonumber\\ - 391259133 h_2^8 + 81882223 h_2^9 - 2063856 h_2^{10} - 
1715472 h_2^{11} + 162944 h_2^{12}\hspace{1.55cm} \nonumber\\- 984855213 h_3^2 + 
6130798389 h_2 h_3^2 - 14569978437 h_2^2 h_3^2+ 
16850898261 h_2^3 h_3^2\hspace{0.85cm} \nonumber\\ - 9439886367 h_2^4 h_3^2 + 
1593033399 h_2^5 h_3^2 + 730653777 h_2^6 h_3^2- 352792017 h_2^7 h_3^2 \hspace{1.1cm}\nonumber\\ + 
42690240 h_2^8 h_3^2 - 562032 h_2^9 h_3^2 + 7812512937 h_3^4 - 
22941081063 h_2 h_3^4 \hspace{2.1cm}\nonumber\\ + 24720233994 h_2^2 h_3^4 - 
11808597150 h_2^3 h_3^4 + 2295385533 h_2^4 h_3^4- 
64422459 h_2^5 h_3^4  \hspace{0.65cm}\nonumber\\- 14031792 h_2^6 h_3^4 - 3311723799 h_3^6 + 
3321565299 h_2 h_3^6 - 982634409 h_2^2 h_3^6 \hspace{1.7cm}\nonumber\\ + 65800269 h_2^3 h_3^6 + 
29760696 h_3^8) \hspace{2.6cm}\nonumber 
\eea
We have calculated also $4$ and $5$ instanton corrections, but the formulae are too lengthy 
to be presented here.
 
By means of numerical integration of the differential equation (\ref{diff_eq2}) along the 
line indicated in section \ref{sect_MM} we have computed the eigenvalues of monodromy matrix
(\ref{M_matrix}) for several values of the instanton parameter $q=\Lambda^6$, namely for 
the values 
\bea
\label{Lambda_range}
\Lambda=\exp \left(\frac{k-1}{20}-5\right) \,,\qquad k=1,2,\cdots,120\,,
\eea  
and fixed values of parameters 
\[
p_1=0.12\,; \qquad p_2=0.17\,; \qquad p_3=-0.29\,.
\]
Due to identification (\ref{spec_M}) this allows to find the corresponding 
$A$-cycle periods $a_1,a_2,a_3$. In table \ref{tab:table1} we present some characteristic 
excerpt from the resulting data.  
\begin{table}[h!]
	\begin{center}
		\scalebox{0.9}{
\begin{tabular}{l|l|l} 
\textbf{$\Lambda$} & $a_1$ & $a_2$\\
\hline
0.00822974704902 & 0.1200000000131 & 0.169999999982\\
0.0223707718562 & 0.1200000053049 & 0.169999992932 \\
0.0608100626252 & 0.1200021402877 & 0.169997148430\\
0.165298888222 & 0.1208841761521 & 0.168828966405\\
0.246596963942 & 0.1349151981823 & 0.151933010167 \\
0.272531793034 & 0.142136769453 - 0.019455438633 i & 0.142136769453 + 0.019455438633 i\\
0.449328964117 & 0.092117229441 - 0.135924390553 i & 0.092117229441 + 0.135924390553 i\\
0.740818220682 & 0.003727137475 - 0.568756791077 i & 0.003727137475 + 0.568756791077 i\\
1.22140275816 & 0.000899023180 - 1.071594057757 i & 0.000899023180 + 1.071594057757 i\\
2.01375270747 & 0.00036203460 - 1.78605985179 i & 0.00036203460 + 1.78605985179 i\\
3.32011692274 & 0.00013130957 - 2.96965962318 & 0.00013132399 + 2.96965962932 i
\end{tabular}}
\end{center}
\caption{The values $a_1$, $a_2$ obtained through numerical integration 
of the differential equation (\ref{diff_eq2}) with initial conditions (\ref{BC}) 
for $p_1=0.12$, $p_2=0.28$. }
\label{tab:table1}
\end{table}  

Inserting the values of $a_k$, $\Lambda$ in (\ref{tr_phi2}), (\ref{tr_phi3})
supplemented by $q^4$ and $q^5$ corrections we have calculated $\langle \textbf{tr}\,\phi^2\rangle$ and 
$\langle \textbf{tr}\,\phi^3\rangle$. The consistency requires that 
at small values of $q$ for at which instanton expansion is valid one should always obtain the same expectation values $\langle \textbf{tr}\,\phi^2\rangle 
=p_1^2+p_2^2+p_3^2=0.1274$ and $\langle \textbf{tr}\,\phi^3\rangle 
=p_1^3+p_2^3+p_3^3=-0.017748$. Table \ref{tab:table2} displays the results of actual
computations.   
\begin{table}[h!]
	\begin{center}
		\begin{tabular}{l|l|l} 
			\textbf{$\Lambda$} & $\langle \textbf{tr}\,\phi^2\rangle$
			 & $\langle \textbf{tr}\,\phi^3\rangle$\\
			\hline
			0.00822974704902 & 0.1274000000000 & -0.0177480000000 \\
			0.0223707718562 & 0.1274000000000 & -0.0177480000000 \\
			0.0608100626252 & 0.1274000000000 & -0.0177480000000 \\
			0.165298888222 & 0.1274000000000 & -0.0177480000000 \\
			0.246596963942 & 0.1273999999998 & -0.0177480000000 \\
			0.272531793034 & 0.1273999999922 & -0.0177479999994 \\
			0.449328964117 & 0.1273774046391 & -0.0177462190257 \\
			0.740818220682 & 0.1313057536866 & -0.0178774876030
		\end{tabular}
	\end{center}
	\caption{The values $\langle \textbf{tr}\,\phi^2\rangle$, 
	$\langle \textbf{tr}\,\phi^3\rangle$  obtained by inserting  
	the values of $a_1$, $a_2$ from Table \ref{tab:table1} into (\ref{tr_phi2}), 
	(\ref{tr_phi3}) supplemented by $q^4$ and $q^5$ corrections. 
	To be compared with (by definition) $\langle \textbf{tr}\,\phi^2\rangle 
	=p_1^2+p_2^2+p_3^2=0.1274$ and $\langle \textbf{tr}\,\phi^3\rangle 
	=p_1^3+p_2^3+p_3^3=-0.017748$.}
	\label{tab:table2}
\end{table} 
One expects an essential deviation from the instanton series starting from 
the value of $\Lambda$ at which the polynomial 
\[
\left(z^3-\frac{u_2}{2}\,z-\frac{u_3}{3}\right)^2-4 \Lambda^6
\]
acquires coinciding zeros, i.e. at the point where its discriminant
\[
1024 \Lambda ^{18} \left(216 \Lambda ^6-72 \Lambda ^3 u_3-u_2^3+6 u_3^2\right) 
\left(216 \Lambda ^6+72 \Lambda ^3 u_3-u_2^3+6 u_3^2\right)
\]
vanishes. Such points correspond to massless dyons or monopoles. It is easy to 
check that within the range of $\Lambda$ 
(\ref{Lambda_range}) the only zero is at $\Lambda=0.1822359934629\cdots $ for which
the last factor of discriminant vanishes. And in fact, inspecting table \ref{tab:table2} 
one sees that for the greater values of $\Lambda$'s, the mismatch becomes significant 
while for smaller values the agreement is quite impressive. 
Notice also from table \ref{tab:table1} that for $\Lambda>0.24659696394 $ 
we encountered complex values for $a_{1}$ and $a_{2}$. 
\subsection{Extension of Zamolodcikov's conjecture to $SU(3)$}
The simpler case of the gauge group $SU(2)$ has been analyzed recently in 
\cite{Fioravanti:2019vxi}. In this case one has to deal with the Mathieu equation. Corresponding $TQ$ relation was investigated in \cite{Zamolodchikov:2000unpb}, where Al. Zamolodchikov conjectured (and demonstrated numerically) an elegant relationship 
between $T$-function and Floquet exponent $\nu$ of Mathieu equation:
\bea 
T=\cosh (2\pi \nu) \, .
\eea
Here we suggest a natural extension of Zamolodchikov's conjecture for $SU(3)$ case:
\bea
\label{conj_su3}
T(\theta)=\sum_{n=1}^3e^{2 \pi i a_n} \, .
\eea 
Notice, that at $\theta\ll 0$ the asymptotic (\ref{Q_asymp}) leads to
\bea 
\label{T_asym}
T(\theta)\sim\sum_{n=1}^3e^{2 \pi i p_n}  \,\, ,
\eea 
which is consistent with (\ref{conj_su3}), since for $\theta\ll 0$ instanton 
corrections disappear and $a_k$ coincides with $p_k$.
\section{Few perspectives}
It would be very interesting to have a TBA for our case and check our 
conjecture (\ref{conj_su3}) as it was done by Al. Zamolodchikov in 
\cite{Zamolodchikov:2000unpb}. Actually, even relevant would be a gauge TBA that may shed light on the dual $B$-cycle periods 
$\bf{a_D}$ along the route presented in \cite{Fioravanti:2019vxi} for the $SU(2)$ case (see also the presentation of \cite{Gaiotto:2014bza} and \cite{Grassi:2019coc}).

It is well known that  pure ${\cal N}=2$ $SU(N)$ theories with $N>2$ are endowed with special points 
in there moduli spaces of vacua at which   mutually nonlocal dyons become massless (Argyres-Douglas points) \cite{ Argyres:1995jj}.
It would be interesting to investigate this phenomenon within our approach (for the NS regime of the $\Omega$ background).

Furthermore, for generic groups of gauge theories (starting with $SU(2)$ and general Liouville ODE/IM correspondence) it is very intriguing to investigate the form of 'potentials' of the ODE describing excited states of the IM ({\it cf.} \cite{BLZ-excited-ode-im, DF-excited-ode-im} for what we know about 'ordinary' ODE/IM). In fact, the latter should be obtainable also via analytic continuation in the parameters/moduli, and thus, these non-trivial monodromies would be of great interest in gauge theories.

Of course, it is very plausible that the imaginable generalizations of our results, and in particular of (\ref{conj_su3}), might hold for arbitrary $SU(N)$ gauge groups. In fact, for those higher order differential equations we shall have also the enlightening treatment of a mathematically similar problem with only one irregular singularity (at $\infty$), the case of gluon scattering amplitudes/Wilson loops at strong coupling in planar ${\cal N}= 4$ SYM \cite{TBA-amp1,TBA-amp3}. Because of its different physical nature, this problem allow for a beautiful and all-coupling exact Operator Product Expansion \cite{TBA-amp2,BSV1}, whose strong coupling limit reproduces interestingly the integrable TBA \cite{FPR} of \cite{TBA-amp1,TBA-amp3}: the similar mathematical structures and ideas of these two different fields should bear fruit, in future, for a deeper understanding.

\section*{Acknowledgments}
HP and RP are grateful to G. Sarkissian and R. Mkrtchyan for many useful discussions.  
DF would like to thank D. Gregori, M. Rossi, R. Tateo for stimulating discussions. This work has been partially supported by the grants: research project 18T-1C340 from Armenian State Committee of Science, GAST (INFN), the MPNS-COST Action MP1210, the EC Network Gatis and the MIUR-PRIN contract 2017CC72MK$\textunderscore$003. D.F. thanks the GGI for Theoretical Physics for invitation to the workshop 'Supersymmetric Quantum Field Theories in the Non-perturbative Regime'.

\begin{appendix}
\section{Proving the $TQ$ relations}
\label{TQ_proof}
In this appendix we prove that different choices of indices $j\ne k $ in 
$TQ$ relations (\ref{T_Q}) are consistent with $QQ$ relations  (\ref{Qc_QQ}). 
 
For example let us choose $j=1$, $k=2$ 
\bea
\label{T_Q_12}
&T(\theta)Q_1\left(\theta-\frac{\pi i}{6}\right)\bar{Q}_2\left(\theta+\frac{\pi i}{6}\right)  
=\\
&Q_1(\theta-\frac{5\pi i}{6})\bar{Q}_2\left(\theta+\frac{\pi i}{6}\right)
+Q_1\left(\theta+\frac{\pi i}{2}\right)\bar{Q}_2\left(\theta-\frac{\pi i}{2}\right)+
Q_1\left(\theta-\frac{\pi i}{6}\right)\bar{Q}_2\left(\theta+\frac{5\pi i}{6}\right),\nonumber  
\eea 
and  $j=1$, $k=3$  
\bea
\label{T_Q_13}
&T(\theta)Q_1\left(\theta-\frac{\pi i}{6}\right)\bar{Q}_3\left(\theta+\frac{\pi i}{6}\right)  
=\\
&Q_1(\theta-\frac{5\pi i}{6})\bar{Q}_3\left(\theta+\frac{\pi i}{6}\right)
+Q_1\left(\theta+\frac{\pi i}{2}\right)\bar{Q}_3\left(\theta-\frac{\pi i}{2}\right)+
Q_1\left(\theta-\frac{\pi i}{6}\right)\bar{Q}_3\left(\theta+\frac{5\pi i}{6}\right),\nonumber  
\eea 
Multiplying (\ref{T_Q_12}) by $\bar{Q}_3\left(\theta+\frac{\pi i}{6}\right) $, 
(\ref{T_Q_13}) by $\bar{Q}_2\left(\theta+\frac{\pi i}{6}\right)$ 
and taking difference, the right hand side becomes
\bea
&\bar{Q}_3\left(\theta+\frac{\pi i}{6}\right)\left(Q_1\left(\theta+\frac{\pi i}{2}\right)\bar{Q}_2\left(\theta-\frac{\pi i}{2}\right)+
Q_1\left(\theta-\frac{\pi i}{6}\right)\bar{Q}_2\left(\theta+\frac{5\pi i}{6}\right)\right)-
\nonumber\\
&\bar{Q}_2\left(\theta+\frac{\pi i}{6}\right)\left(
Q_1\left(\theta+\frac{\pi i}{2}\right)\bar{Q}_3\left(\theta-\frac{\pi i}{2}\right)+
Q_1\left(\theta-\frac{\pi i}{6}\right)\bar{Q}_3\left(\theta+\frac{5\pi i}{6}\right)
\right)=\\
&Q_1\left(\theta+\frac{\pi i}{2}\right)
\left(
\bar{Q}_3\left(\theta+\frac{\pi i}{6}\right)\bar{Q}_2\left(\theta-\frac{\pi i}{2}\right)-
\bar{Q}_2\left(\theta+\frac{\pi i}{6}\right)\bar{Q}_3\left(\theta-\frac{\pi i}{2}\right)\right)+
\nonumber\\
&Q_1\left(\theta-\frac{\pi i}{6}\right)\left(
\bar{Q}_3\left(\theta+\frac{\pi i}{6}\right)\bar{Q}_2\left(\theta+\frac{5\pi i}{6}\right)-
\bar{Q}_2\left(\theta+\frac{\pi i}{6}\right)\bar{Q}_3\left(\theta+\frac{5\pi i}{6}\right)
\right)\,.
\nonumber
\eea
Obviously the last expression vanishes due to (\ref{Qc_QQ}) as consistency requires.

Now let us show that $T(\theta)$ does not have any pole. 
It follows from (\ref{T_Q_12}), that a potential pole of $T(\theta)$ can be found 
either among the zeros of $Q_1\left(\theta-\frac{\pi i}{6}\right)$ or 
$\bar{Q}_2\left(\theta+\frac{\pi i}{6}\right)$. For definiteness let us assume that 
it belongs to zero set of  $Q_1\left(\theta-\frac{\pi i}{6}\right)$ (the other option 
can be considered in completely analogues manner). Let $Q_1(\theta_\ell)=0$. 
Then for $\theta=\theta_\ell+\frac{i\pi}{6}$ the r.h.s. of (\ref{T_Q_12}) is equal to 
\bea
\label{T_entire}
&Q_1(\theta_\ell-\frac{2\pi i}{3})\bar{Q}_2\left(\theta_\ell+\frac{\pi i}{3}\right)
+Q_1\left(\theta_\ell+\frac{2\pi i}{3}\right)\bar{Q}_2\left(\theta_\ell-\frac{\pi i}{3}\right)=\\
&\frac{2i \pi^2}{\sin(\pi p_{31})}\left(Q_1(\theta_\ell-\frac{2\pi i}{3})Q_1\left(\theta_\ell+\frac{2\pi i}{3}\right)Q_3\left(\theta_\ell\right)-
Q_1\left(\theta_\ell+\frac{2\pi i}{3}\right)Q_1(\theta_\ell-\frac{2\pi i}{3})Q_3\left(\theta_\ell\right)\right)=0,
\nonumber
\eea
where the relations  (\ref{Qc_QQ}) for both 
$\bar{Q}_2\left(\theta_\ell\pm\frac{\pi i}{3}\right)$ has been used. So $T(\theta_\ell)$ is finite, 
thus proving that $T(\theta)$ is entire.
\end{appendix}
\bibliographystyle{JHEP}

\begin{thebibliography}{10}

\bibitem{Seiberg:1994rs}
N.~Seiberg and E.~Witten, {\it {Electric - magnetic duality, monopole
		condensation, and confinement in N=2 supersymmetric Yang-Mills theory}},
{\em Nucl. Phys.} {\bf B426} (1994) 19--52,
[\href{http://arxiv.org/abs/hep-th/9407087}{{\tt hep-th/9407087}}]. [Erratum:
Nucl. Phys.B430,485(1994)].
	
	\bibitem{Dorey:2002ik}
	N.~Dorey, T.~J. Hollowood, V.~V. Khoze, and M.~P. Mattis, {\it {The Calculus of
			many instantons}},  {\em Phys. Rept.} {\bf 371} (2002) 231--459,
	[\href{http://arxiv.org/abs/hep-th/0206063}{{\tt hep-th/0206063}}].
	
	\bibitem{Flume:2001nc}
	R.~Flume, R.~Poghossian, and H.~Storch, {\it {The Coefficients of the
			Seiberg-Witten prepotential as intersection numbers(?)}},
	\href{http://arxiv.org/abs/hep-th/0110240}{{\tt hep-th/0110240}}.
	
	\bibitem{Flume:2001kb}
	R.~Flume, R.~Poghossian, and H.~Storch, {\it {The Seiberg-Witten prepotential
			and the Euler class of the reduced moduli space of instantons}},  {\em Mod.
		Phys. Lett.} {\bf A17} (2002) 327--340,
	[\href{http://arxiv.org/abs/hep-th/0112211}{{\tt hep-th/0112211}}].
	
	\bibitem{Nekrasov:2002qd}
	N.~A. Nekrasov, {\it {Seiberg-Witten prepotential from instanton counting}},
	{\em Adv. Theor. Math. Phys.} {\bf 7} (2003), no.~5 831--864,
	[\href{http://arxiv.org/abs/hep-th/0206161}{{\tt hep-th/0206161}}].
	
	\bibitem{Flume:2002az}
	R.~Flume and R.~Poghossian, {\it {An Algorithm for the microscopic evaluation
			of the coefficients of the Seiberg-Witten prepotential}},  {\em Int. J. Mod.
		Phys.} {\bf A18} (2003) 2541,
	[\href{http://arxiv.org/abs/hep-th/0208176}{{\tt hep-th/0208176}}].
	
	\bibitem{Nekrasov:2003rj}
	N.~Nekrasov and A.~Okounkov, {\it {Seiberg-Witten theory and random
			partitions}},  {\em Prog. Math.} {\bf 244} (2006) 525--596,
	[\href{http://arxiv.org/abs/hep-th/0306238}{{\tt hep-th/0306238}}].
	
	\bibitem{Bruzzo:2002xf}
	U.~Bruzzo, F.~Fucito, J.~F. Morales, and A.~Tanzini, {\it {Multiinstanton
			calculus and equivariant cohomology}},  {\em JHEP} {\bf 05} (2003) 054,
	[\href{http://arxiv.org/abs/hep-th/0211108}{{\tt hep-th/0211108}}].
	
	\bibitem{Alday:2009aq}
	L.~F. Alday, D.~Gaiotto, and Y.~Tachikawa, {\it {Liouville Correlation
			Functions from Four-dimensional Gauge Theories}},  {\em Lett. Math. Phys.}
	{\bf 91} (2010) 167--197, [\href{http://arxiv.org/abs/0906.3219}{{\tt
			arXiv:0906.3219}}].
	
	\bibitem{Nekrasov:2009rc}
	N.~A. Nekrasov and S.~L. Shatashvili, {\it {Quantization of Integrable Systems
			and Four Dimensional Gauge Theories}},  in {\em {Proceedings, 16th
			International Congress on Mathematical Physics (ICMP09)}}, 2009.
	\newblock \href{http://arxiv.org/abs/0908.4052}{{\tt arXiv:0908.4052}}.
	
	\bibitem{Poghossian:2010pn}
	R.~Poghossian, {\it {Deforming SW curve}},  {\em JHEP} {\bf 04} (2011) 033,
	[\href{http://arxiv.org/abs/1006.4822}{{\tt arXiv:1006.4822}}].
	
	\bibitem{Fucito:2011pn}
	F.~Fucito, J.~F. Morales, D.~R. Pacifici, and R.~Poghossian, {\it {Gauge
			theories on $\Omega$-backgrounds from non commutative Seiberg-Witten
			curves}},  {\em JHEP} {\bf 05} (2011) 098,
	[\href{http://arxiv.org/abs/1103.4495}{{\tt arXiv:1103.4495}}].
	
	
	\bibitem{Bourgine:2017aqk}
	J.-E. Bourgine and D.~Fioravanti, ``{Quantum integrability of $\mathcal{N}=2$
  4d gauge theories},'' \href{http://dx.doi.org/10.1007/JHEP08(2018)125}{{\em
  JHEP} {\bfseries 08} (2018) 125},
  \href{http://arxiv.org/abs/1711.07935}{{\ttfamily arXiv:1711.07935
  [hep-th]}}.
  		
	
	
	\bibitem{Belavin:1984vu}
	A.~A. Belavin, A.~M. Polyakov, and A.~B. Zamolodchikov, {\it {Infinite
			Conformal Symmetry in Two-Dimensional Quantum Field Theory}},  {\em Nucl.
		Phys.} {\bf B241} (1984) 333--380.
	
	\bibitem{Alday:2009fs}
	L.~F. Alday, D.~Gaiotto, S.~Gukov, Y.~Tachikawa, and H.~Verlinde, {\it {Loop
			and surface operators in N=2 gauge theory and Liouville modular geometry}},
	{\em JHEP} {\bf 01} (2010) 113, [\href{http://arxiv.org/abs/0909.0945}{{\tt
			arXiv:0909.0945}}].
	
	\bibitem{Mironov:2009uv}
	A.~Mironov and A.~Morozov, {\it {Nekrasov Functions and Exact Bohr-Zommerfeld
			Integrals}},  {\em JHEP} {\bf 04} (2010) 040,
	[\href{http://arxiv.org/abs/0910.5670}{{\tt arXiv:0910.5670}}].
	
	\bibitem{Maruyoshi:2010iu}
	K.~Maruyoshi and M.~Taki, {\it {Deformed Prepotential, Quantum Integrable
			System and Liouville Field Theory}},  {\em Nucl. Phys.} {\bf B841} (2010)
	388--425, [\href{http://arxiv.org/abs/1006.4505}{{\tt arXiv:1006.4505}}].
	
	\bibitem{Marshakov:2010fx}
	A.~Marshakov, A.~Mironov, and A.~Morozov, {\it {On AGT Relations with Surface
			Operator Insertion and Stationary Limit of Beta-Ensembles}},  {\em J. Geom.
		Phys.} {\bf 61} (2011) 1203--1222,
	[\href{http://arxiv.org/abs/1011.4491}{{\tt arXiv:1011.4491}}].
	
	\bibitem{Nekrasov:2013xda}
	N.~Nekrasov, V.~Pestun, and S.~Shatashvili, {\it {Quantum geometry and quiver
			gauge theories}},  \href{http://arxiv.org/abs/1312.6689}{{\tt
			arXiv:1312.6689}}.
	
	\bibitem{Piatek:2011tp}
	M.~Piatek, {\it {Classical conformal blocks from TBA for the elliptic
			Calogero-Moser system}},  {\em JHEP} {\bf 06} (2011) 050,
	[\href{http://arxiv.org/abs/1102.5403}{{\tt arXiv:1102.5403}}].
	
	\bibitem{Ashok:2015gfa}
	S.~K. Ashok, M.~Bill\'o, E.~Dell'Aquila, M.~Frau, R.~R. John, and A.~Lerda,
	{\it {Non-perturbative studies of N=2 conformal quiver gauge theories}},
	{\em Fortsch. Phys.} {\bf 63} (2015) 259--293,
	[\href{http://arxiv.org/abs/1502.05581}{{\tt arXiv:1502.05581}}].
	
	\bibitem{Poghossian:2016rzb}
	R.~Poghossian, {\it {Deformed SW curve and the null vector decoupling equation
			in Toda field theory}},  \href{http://arxiv.org/abs/1601.05096}{{\tt
			arXiv:1601.05096}}.
	
	\bibitem{Poghosyan:2016mkh}
	G.~Poghosyan and R.~Poghossian, {\it {VEV of Baxter's Q-operator in N=2 gauge
			theory and the BPZ differential equation}},  {\em JHEP} {\bf 11} (2016) 058,
	[\href{http://arxiv.org/abs/1602.02772}{{\tt arXiv:1602.02772}}].
	
	\bibitem{Nekrasov:2017gzb}
	N.~Nekrasov, {\it {BPS/CFT correspondence V: BPZ and KZ equations from
			qq-characters}},  \href{http://arxiv.org/abs/1711.11582}{{\tt
			arXiv:1711.11582}}.
	
	\bibitem{Jeong:2017mfh}
	S.~Jeong and X.~Zhang, {\it {BPZ equations for higher degenerate fields and
			non-perturbative Dyson-Schwinger equations}},
	\href{http://arxiv.org/abs/1710.06970}{{\tt arXiv:1710.06970}}.
	
	
	
\bibitem{Gaiotto:2009we}
  D.~Gaiotto,
  {\it {N=2 dualities}},
  {\em JHEP} {\bf 1208} (2012) 034,
  \href{http://arxiv.org/abs/0904.2715}{{\tt arXiv:0904.2715}}.
	
  
  
  
	
	\bibitem{Dorey:1998pt}
	P.~Dorey and R.~Tateo, {\it {Anharmonic oscillators, the thermodynamic Bethe
			ansatz, and nonlinear integral equations}},  {\em J. Phys.} {\bf A32} (1999)
	L419--L425, [\href{http://arxiv.org/abs/hep-th/9812211}{{\tt
			hep-th/9812211}}].
	
	\bibitem{Bazhanov:1998wj}
	V.~V. Bazhanov, S.~L. Lukyanov, and A.~B. Zamolodchikov, {\it {Spectral
			determinants for Schrodinger equation and Q operators of conformal field
			theory}},  {\em J. Statist. Phys.} {\bf 102} (2001) 567--576,
	[\href{http://arxiv.org/abs/hep-th/9812247}{{\tt hep-th/9812247}}].
	
	
	
	\bibitem{Fioravanti:2019vxi}
	D.~Fioravanti and D.~Gregori, {\it {Integrability and cycles of deformed ${\cal
				N}=2$ gauge theory}},  \href{http://arxiv.org/abs/1908.08030}{{\tt
			arXiv:1908.08030}}.



	
	\bibitem{Gaiotto:2014bza}
	D.~Gaiotto, {\it {Opers and TBA}},  \href{http://arxiv.org/abs/1403.6137}{{\tt
			arXiv:1403.6137}}.
	
	\bibitem{Grassi:2019coc}
	A.~Grassi, J.~Gu, and M.~Mari\~{n}o, {\it {Non-perturbative approaches to the
			quantum Seiberg-Witten curve}},  \href{http://arxiv.org/abs/1908.07065}{{\tt
			arXiv:1908.07065}}.



\bibitem{Klemm:1994qs}
  A.~Klemm, W.~Lerche, S.~Yankielowicz and S.~Theisen,
  {\it Simple singularities and N=2 supersymmetric Yang-Mills theory},
  Phys.\ Lett.\ B {\bf 344} (1995) 169,
 \href{https://arxiv.org/abs/hep-th/9411048}{{\ttfamily arXiv:hep-th/9411048}}
 
 
   
  	

\bibitem{Argyres:1994xh} 
P.~C.~Argyres and A.~E.~Faraggi,
{\it The vacuum structure and spectrum of N=2 supersymmetric SU(n) gauge theory},
Phys.\ Rev.\ Lett.\  {\bf 74},  (1995) 3931,
\href{https://arxiv.org/abs/hep-th/9411057}{{\ttfamily arXiv:hep-th/9411057}}.



	
	
	
	\bibitem{Dorey:1999pv}
	P.~Dorey and R.~Tateo, {\it {Differential equations and integrable models: The
			SU(3) case}},  {\em Nucl. Phys.} {\bf B571} (2000) 583--606,
	[\href{http://arxiv.org/abs/hep-th/9910102}{{\tt hep-th/9910102}}]. [Erratum:
	Nucl. Phys.B603,582(2001)].
	
	\bibitem{Bazhanov:2001xm}
	V.~V. Bazhanov, A.~N. Hibberd, and S.~M. Khoroshkin, {\it {Integrable structure
			of W(3) conformal field theory, quantum Boussinesq theory and boundary affine
			Toda theory}},  {\em Nucl. Phys.} {\bf B622} (2002) 475--547,
	[\href{http://arxiv.org/abs/hep-th/0105177}{{\tt hep-th/0105177}}].
	
	
		
				
		\bibitem{Dorey:1999uk}
  P.~Dorey and R.~Tateo,
  ``On the relation between Stokes multipliers and the T-Q systems of conformal field theory,''
  Nucl.\ Phys.\ B {\bf 563} (1999) 573
   Erratum: [Nucl.\ Phys.\ B {\bf 603} (2001) 581]
 and 
 \href{https://arxiv.org/abs/hep-th/9906219}{{\ttfamily arXiv:hep-th/9906219}}
			
	
	\bibitem{Zamolodchikov:2000unpb}
	Al.B.~Zamolodchikov, {\it {Generalized Mathieu equation and Liouville TBA, 2000}},
	{\em in Quantum Field Theories in Two Dimensions, vol. 2, World Scientific,
		2012}.
	
	\bibitem{Losev:2003py}
	A.~S. Losev, A.~Marshakov, and N.~A. Nekrasov, {\it {Small instantons, little
			strings and free fermions}},  \href{http://arxiv.org/abs/hep-th/0302191}{{\tt
			hep-th/0302191}}.
	
	\bibitem{Flume:2004rp}
	R.~Flume, F.~Fucito, J.~F. Morales, and R.~Poghossian, {\it {Matone's relation
			in the presence of gravitational couplings}},  {\em JHEP} {\bf 04} (2004)
	008, [\href{http://arxiv.org/abs/hep-th/0403057}{{\tt hep-th/0403057}}].
	
	\bibitem{NIST:DLMF}
	``{\it NIST Digital Library of Mathematical Functions}.''
	http://dlmf.nist.gov/16\\Chapter 16, A.~B.~Olde Daalhuis, R.~A.~Askey,
	Generalized Hypergeometric Functions and Meijer G-Function.
	
	\bibitem{Matone:1995rx}
	M.~Matone, {\it {Instantons and recursion relations in N=2 SUSY gauge theory}},
	{\em Phys. Lett.} {\bf B357} (1995) 342--348,
	[\href{http://arxiv.org/abs/hep-th/9506102}{{\tt hep-th/9506102}}].
	
	
	


\bibitem{Argyres:1995jj} 
P.~C.~Argyres and M.~R.~Douglas,
{\it New phenomena in SU(3) supersymmetric gauge theory},
Nucl.\ Phys.\ B {\bf 448}, 93 (1995)
doi:10.1016/0550-3213(95)00281-V
and
\href{https://arxiv.org/abs/hep-th/9505062}{{\ttfamily arXiv:hep-th/9505062}}.

	
	
	\bibitem{BLZ-excited-ode-im}
V.~Bazhanov, S.~Lukyanov, and A.~Zamolodchikov, ``{Higher level
  eigenvalues of Q operators and Schroedinger equation},''
  \href{http://dx.doi.org/10.4310/ATMP.2003.v7.n4.a4}{{\em Adv. Theor. Math.
  Phys.} {\bfseries 7} no.~4, (2003) 711--725},
\href{http://arxiv.org/abs/hep-th/0307108}{{\ttfamily arXiv:hep-th/0307108
  [hep-th]}}.


\bibitem{DF-excited-ode-im}
D.~Fioravanti, ``{Geometrical loci and CFTs via the Virasoro symmetry of the
  mKdV-SG hierarchy: An Excursus},''
  \href{http://dx.doi.org/10.1016/j.physletb.2005.01.037}{{\em Phys. Lett.}
  {\bfseries B609} (2005) 173--179},
\href{http://arxiv.org/abs/hep-th/0408079}{{\ttfamily arXiv:hep-th/0408079
  [hep-th]}}.
  
  
  
   \bibitem{TBA-amp1}
L.F. Alday, D. Gaiotto, J.M. Maldacena,
{\sl Thermodynamic Bubble Ansatz},
JHEP{\bf 09} (2011) 032
and
\href{https://arxiv.org/abs/0911.4708}{{\ttfamily arXiv:hep-th/0911.4708}}

\bibitem{TBA-amp3}
L.F. Alday, J.M. Maldacena, A. Sever, P. Vieira
{\sl Y-system for Scattering Amplitudes},
J.Phys. {\bf A43} (2010) 485401 and
\href{https://arxiv.org/abs/1002.2459}{{\ttfamily arXiv:hep-th/1002.2459}}

\bibitem{TBA-amp2}
L.F. Alday, D. Gaiotto, J.M. Maldacena, A. Sever, P. Vieira,
{\sl An Operator Product Expansion for Polygonal null Wilson Loops},
JHEP{\bf 04} (2011) 088
and 
\href{https://arxiv.org/abs/1006.2788}{{\ttfamily arXiv:hep-th/1006.2788}}


\bibitem{BSV1}
B. Basso, A. Sever, P. Vieira,
{\sl Space-time S-matrix and Flux-tube S-matrix at Finite Coupling},
Phys. Rev. Lett. {\bf 111} (2013) 091602 and
\href{https://arxiv.org/abs/1303.1396}{{\ttfamily arXiv:hep-th/1303.1396}}



\bibitem{FPR}
D. Fioravanti, S. Piscaglia, M. Rossi,
{\sl Asymptotic Bethe Ansatz on the GKP vacuum as a defect spin chain: scattering, particles and minimal area Wilson loops},
Nucl. Phys. {\bf B898} (2015) 301 and
\href{https://arxiv.org/abs/1503.08795}{{\ttfamily arXiv:hep-th/1503.08795}}


	
	
	
	
\end{thebibliography}
\providecommand{\href}[2]{#2}
\begingroup\raggedright
\endgroup
	
\end{document}